\documentclass[longauth,traditabstract]{aa} 
\usepackage{graphicx}
\usepackage{natbib}
\bibpunct{(}{)}{;}{a}{}{,}
\usepackage{txfonts}
\begin{document}
\title{ Removing systematics from the CoRoT 
\thanks{ The CoRoT space mission, launched on December 27th 2006, has
been developed and is operated by CNES, with the contribution of
Austria, Belgium, Brazil, ESA, Germany, and Spain. CoRoT data become
publicly available one year after release to the Co-Is of the
mission from the CoRoT archive: {\tt
http://idoc-corot.ias.u-psud.fr/}.}  light curves: I.
Magnitude-Dependent Zero Point}

\author{
     T.\ Mazeh \inst{1,2} 
\and P.\ Guterman\inst{3} 
\and S.\ Aigrain \inst{4}
\and S.\ Zucker \inst{5} 
\and N.\ Grinberg\inst{6} \and
A.\ Alapini\inst{4} \and %
     R.\ Alonso\inst{3} \and 
     M.\ Auvergne\inst{7} \and 
     M.\ Barbieri\inst{3} \and 
     P.\ Barge\inst{3} \and 
     P.\ Bord{\' e}\inst{8} \and 
     F.\ Bouchy\inst{9} \and 
     H.\ Deeg\inst{10} \and 
     R.\ De la Reza\inst{11} \and 
     M.\ Deleuil\inst{3} \and 
     R.\ Dvorak\inst{12} \and 
     A.\ Erikson\inst{13} \and 
     M.\ Fridlund\inst{14} \and 
     P.\ Gondoin\inst{14} \and 
     L.\ Jorda\inst{3} \and 
     H.\ Lammer\inst{15} \and 
     A.\ L{\' e}ger\inst{8} \and 
     A.\ Llebaria\inst{3} \and 
     P.\ Magain\inst{16} \and 
     C.\ Moutou\inst{3} \and 
     M.\ Ollivier\inst{8} \and 
     M.\ Paetzold\inst{17} \and 
     F.\ Pont\inst{4} \and 
     D.\ Queloz\inst{18} \and 
     H.\ Rauer\inst{13,20} \and 
     D.\ Rouan\inst{7} \and 
     R.\ Sabo\inst{6} \and
     J.\ Schneider\inst{19} \and 
     G.\ Wuchterl\inst{20} 
}

\institute{ 
Radcliffe Institute for Advanced Studies at Harvard, 8 Garden St.,
Cambridge, MA 02138, USA
\and
On Sabbatical leave from School of Physics and Astronomy, 
    Raymond and Beverly Sackler Faculty of Exact Sciences, 
    Tel Aviv University, Tel Aviv  69978, Israel
\and
Le Laboratoire d'Astrophysique de Marseille, UMR 6110 CNRS, Technop\^ole de Marseille-Etoile, F-13388 Marseille cedex 13, France 
\and
School of Physics, University of Exeter, Physics Building, Stocker
Road, Exeter, Ex4 4QL, UK
\and
Geophysics and Planetary Sciences, Raymond and Beverly Sackler Faculty
of Exact Sciences, Tel Aviv University, Tel Aviv  69978, Israel 
\and
School of Physics and Astronomy, 
Raymond and Beverly Sackler Faculty of Exact Sciences, 
Tel Aviv University, Tel Aviv 69978, Israel
\and
LESIA, Observatoire de Paris, 92195 Meudon, France \and
IAS, Universit{\' e} Paris XI, 91405 Orsay, France \and
Observatoire de Haute-Provence, 04870 St Michel l'Observatoire, France \and 
IAC, E-38205 La Laguna, Spain \and
ON/MCT, 20921-030, Rio de Janeiro, Brazil \and
IfA, University of Vienna, 1180 Vienna, Austria \and
Institute of Planetary Research, DLR, 12489 Berlin, Germany \and
RSSD, ESA/ESTEC, 2200 Noordwijk, The Netherlands \and
IWF, Austrian Academy of Sciences, A-8042 Graz, Austria \and
IAG, Universit{\' e} de Li{\` e}ge, Li{\` e}ge 1, Belgium \and
RIU, Universit{\"a}t zu K{\" o}ln, 50931 K{\" o}ln, Germany \and
Observatoire de Gen{\` e}ve, 1290 Sauverny, Switzerland \and
ZAA, TU Berlin, D-10623 Berlin, Germany \and
LUTH, Observatoire de Paris, 92195 Meudon, France \and
Th{\" u}ringer Landessternwarte, 07778 Tautenburg, Germany 
}
\date{Received / Accepted }

  \abstract{ This paper presents an analysis that searched for
systematic effects within the CoRoT exoplanet field light curves. The
analysis identified a systematic effect that modified the zero point
of most CoRoT exposures as a function of stellar magnitude. We could
find this effect only after preparing a set of learning light
curves that were relatively free of stellar and instrumental noise.
Correcting for this effect, rejecting outliers that appear in almost
every exposure, and applying SysRem, reduced the stellar RMS by about
20\%, without attenuating transit signals.  }

\keywords{methods: data analysis -- methods: statistical --
  techniques: photometric -- planetary systems -- surveys}
   
\titlerunning{Cleaning the CoRoT light curves}	       

\maketitle

%
\section{Introduction}        
\label{introduction}                            
%

CoRoT was launched on December 2006 to perform wide-field accurate
stellar photometry \citep{rouan98, baglin06}, aiming to study the
internal structure of stars using seismology and to detect small
planets using the transit method. Both goals required ultra-high
accuracy, interruption-free photometry, which the space platform, free
from atmospheric disturbances, should afford. Nevertheless, the first
six transiting planets discovered by CoRoT, CoRoT-1b to -6b
\citep[][Rauer et al. submitted; Fridlund et al., in preparation]{barge08,
alonso08, deleuil08, aigrain08}, all have a transit depth of 1--2\%,
easily detectable by ground-based photometry. Only at the first CoRoT
international conference, held in February 2009, the CoRoT team
announced the discovery of the super-Earth CoRoT-7b (Leger et al.,
submitted), with transit depth of 0.03\%, which ground-based
observations could never have detected.

The light curve of the bright star CoRoT-7 is of exceptional quality,
with a RMS scatter over transit timescales of only 0.01\%. However,
the majority of CoRoT light curves contain systematics and correlated
noise, which is probably associated with satellite jitter, stellar
activity, cosmic ray impacts and possibly other effects. Albeit
extremely low compared to ground-based surveys \citep{aigrain09}, this
noise should nonetheless be removed before planets with shallow
transits like the ones of CoRoT-7b could be detected.

In order to remove the systematic effects we have previously applied
SysRem \citep{tmz05} to the CoRoT data, resulting in reduction of the
noise level by 10--20\%. However, SysRem was sensitive to {\it
temporal} variability that was shared by many of the stars, and was
not specifically tuned to detect collective effects that showed up
when considering the measurements of each exposure separately. One
such obvious effect was the zero point of the different exposures,
which could have been modulated by the satellite motion in and out of
the Earth shadow and going through the South Atlantic Anomaly, and
could also depend on the stellar characteristics. When we searched for
such an effect in the LRa01 CoRoT run we discovered that the zero
points of most exposures depended on the stellar magnitude. This could
be noticed only after we prepared a 'learning' set of light curves
that were relatively free of stellar and instrumental noise. When this
effect was removed, the noise level was reduced by about 20\% for the
faint stars.

Section 2 details our analysis, including the preparation of the
learning set of light curves, and the procedure to remove the
effect found, and Section 3 discusses in short our findings.

%
\section{Data analysis}        
\label{analysis}                            
%

Our analysis was done on the white light curves obtained during the
150-day CoRoT LRa01 run \citep{auvergne09}.  We first
transformed the CoRoT fluxes into magnitudes, and then subtracted from
each star its median magnitude.  The residuals of each measurement
relative to its stellar median --- $\{r_{ij}\}$, where $i$ was the
star number and $j$ was the exposure number, were the subject of this
analysis. We concentrated on the residuals derived for any given
exposure $j_0$: $\{r_{ij_0};i=1,M\}$, where $M$ was the number of
stars in this CoRoT run, searching for systematic effects in exposure
$j_0$.

In order to be able to notice relatively small systematic effects we
prepared 'clean' learning light curves by the following measures:

\begin{itemize}
\item
Divided the LRa01 run into 13 blocks, each block contained data taken
over only ten days. 
\item
Removed invalid data marked by the CoRoT regular N2 pipeline
and rejected outliers from each light curve on a temporal identification
basis.
\item
In each block, rejected the stars that showed high variability in that
block, either because of stellar variability or because of obvious
instrumental effects, such as hot pixel features. Thus, some stars appeared
only in some of the blocks.
\item
Removed long-term variability by subtracting a running median, taken
over 3 satellite orbits. Small hot-pixel features were also attenuated
by the median filtering.
\item
Considered the light curves from the two CCDs separately, but combined
the two halves of each CCD together. This was done as we suspected
that the two CCDs will have different systematics.
\item
Some of the light curves were sampled with a rate of 32 sec, while
most stars with a rate of 512 sec. To make the analysis even more
difficult, some stars had a light curve that was first sampled with
the 512 sec rate, which was then switched into the 32 sec. We
therefore had to put all the 512 sec and 32 sec measurements of CoRoT
together in a common synchronous grid.  An accurate common 512 sec
exposure time was emulated as CoRoT would have done on board, rather
than obtained by interpolation. This enabled us to associate each of the
stellar measurements with the correct exposure.
\end{itemize}

As a result of our selection, we were left, in the first block on the
E1 CCD for example, with 4111 learning light curves out of 5704 ones.
Only after the preparation of the 'learning' sets of light curves the
collective features of the data appeared in our analysis with
relatively small spread, so we could accurately estimate and subrtact
these features.

We found that the zero points of most exposures depended on stellar
magnitude, an effect that we dubbed MagZeP (Magnitude Zero Point). To
show this effect we plotted in Figure~1 the residuals of two
exposures, no. 12 and 55, from our first block. The figure showed two
prominent features of the MagZeP effect:
\begin{itemize}
\item
The scatter around the general trend was larger for fainter stars
\item
The residuals took a parabolic shape, with different curvature sign
and slope for different exposures. 
The typical amplitudes of this effect were 0.005 mag for the faint objects. 
\end{itemize}
While the first feature was expected, as the S/N is smaller, the
second effect took us by surprise.


\begin{figure*}
\centering
\resizebox{14cm}{7cm}
{\includegraphics{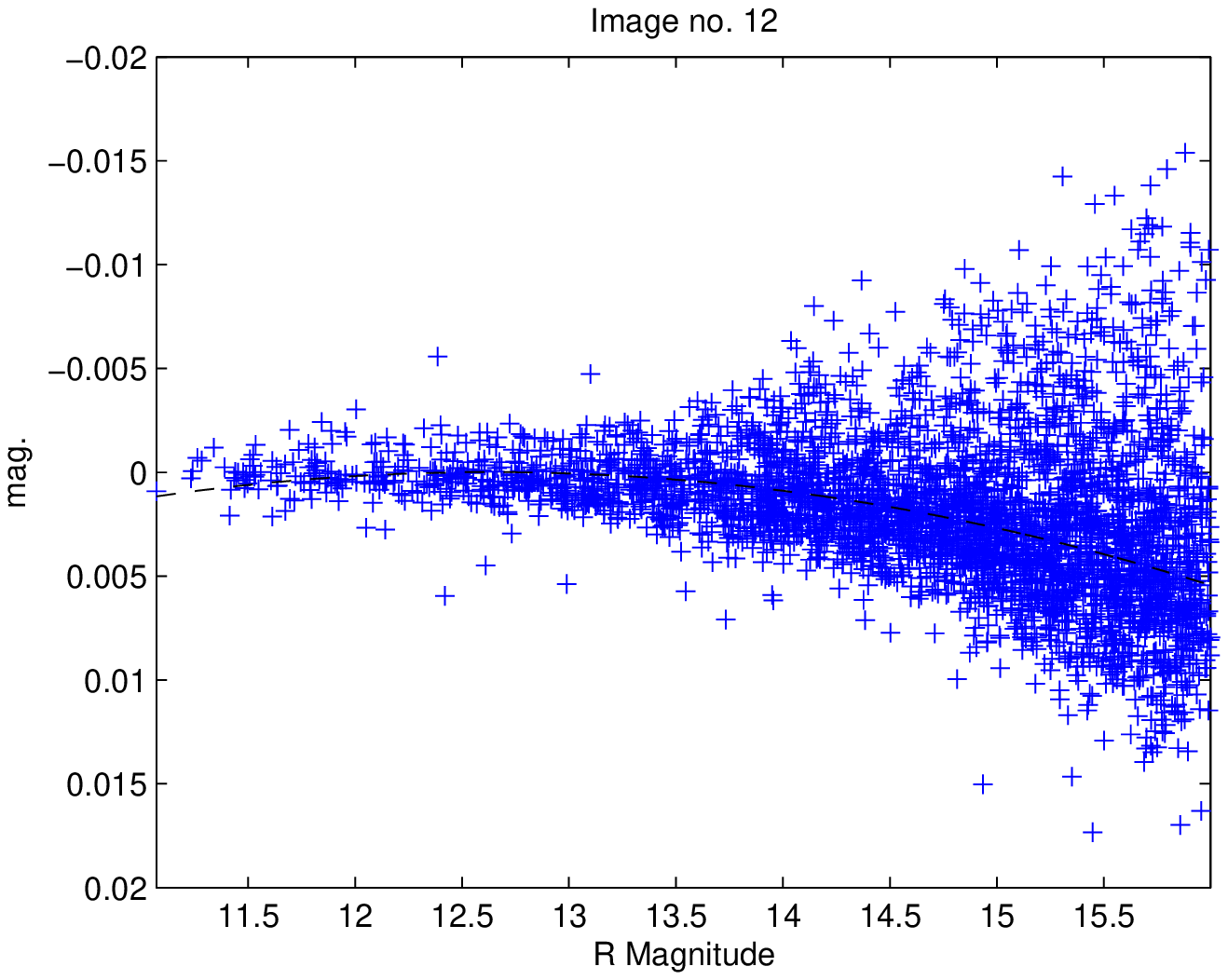}
\includegraphics{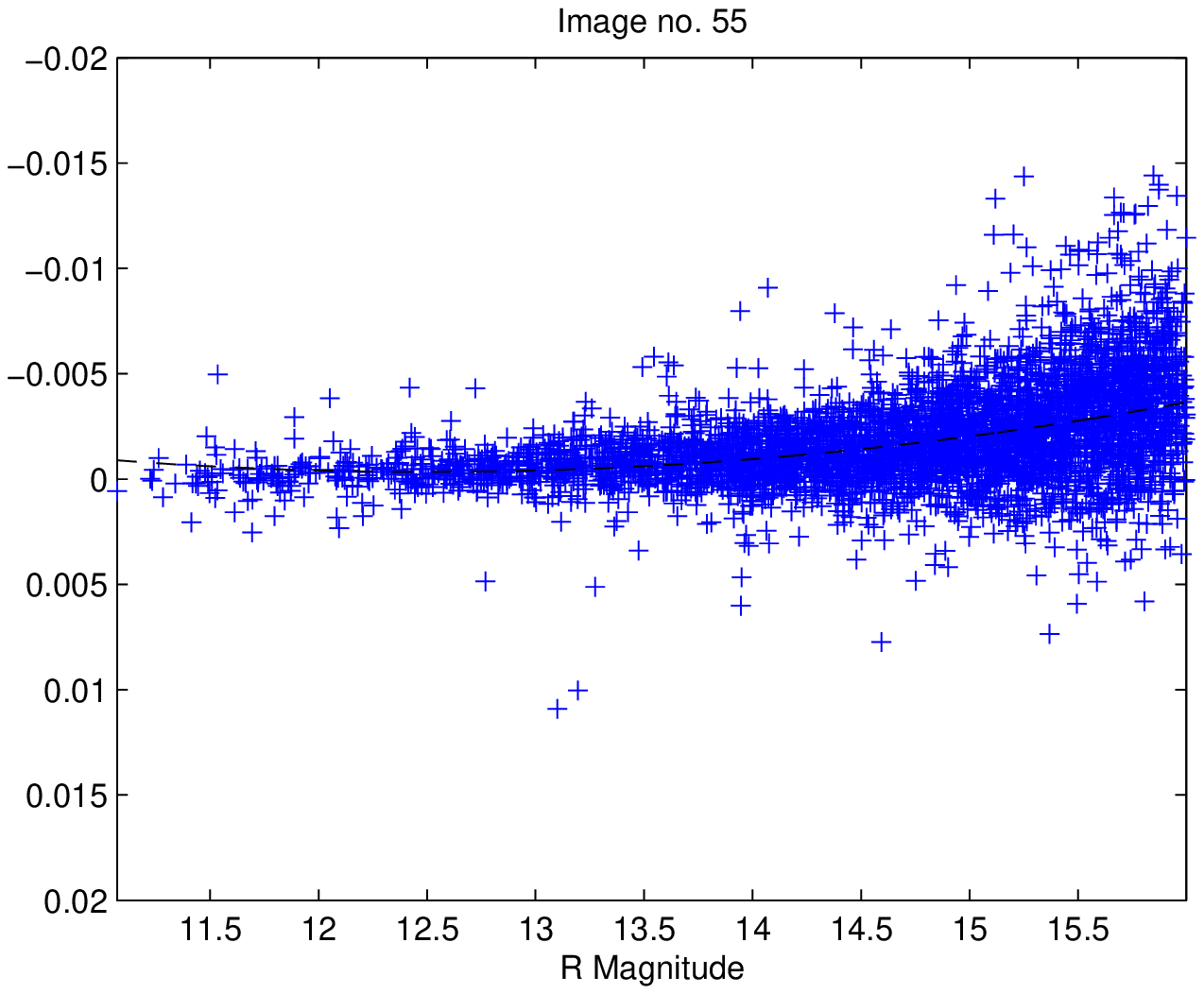}}
\caption{The residuals of two exposures as a function of their stellar
R magnitude. The continuous line is our parabolic fit.  The
left-hand-side panel shows the residuals of exposure
2007-10-24T14:09:15.000, while the right-hand-side panel comes from
exposure 2007-10-24T20:16:11.000.  }
\label{fig1}
\end{figure*}

To remove the MagZeP effect we fitted the residuals of each exposure
$j$ by a parabola that depended on the stellar magnitude $m_i$:

\begin{equation}
p_{ij}=a_{0,j}+a_{1,j}m_i+a_{2,j}m_i^2 \ ,
\end{equation}
and then derived new residuals
\begin{equation}
\tilde r_{ij}=r_{ij}-p_{ij} \ ,
\end{equation}
where $\tilde r_{ij}$ presented our best
estimate of the stellar magnitude, relative to its median.

We also noticed that for almost every exposure some outliers clearly
stood out, as could be seen in Figure 1. To identify these outliers we
assigned to each measurement an error, $\sigma_{ij}$, based on the
collective scatter at the corresponding exposure and magnitude.  We
then rejected residuals that were {\it smaller} than zero (after the
removal of the parabola) by more than $\eta$ times their error
\begin{equation}
 \tilde r_{ij}  \, \leq \eta{\sigma}_{ij}\ ,
\end{equation}
where $\eta$ was a parameter, taken in our present implementation of
the algorithm to be equal to 2. Keeping the positive residuals ensured
us that we did not attenuate any transit signal. Removing these
exposure outliers further improved the stellar scatter. The
parabola was derived by a robust regression based on iteratively
reweighed least-squares fit, which was relatively insensitive to the
presence of outliers. In this way, a robust result was obtained even
without a need for additional fitting-clipping cycle.

Finally, {\it all} CoRoT light curves were cleaned with the parameters
determined by the selected learning set of light curves.  This step of
the analysis produced a homogeneous set of light curves in each block.


\begin{figure*}
\centering \resizebox{10cm}{8cm} {\includegraphics{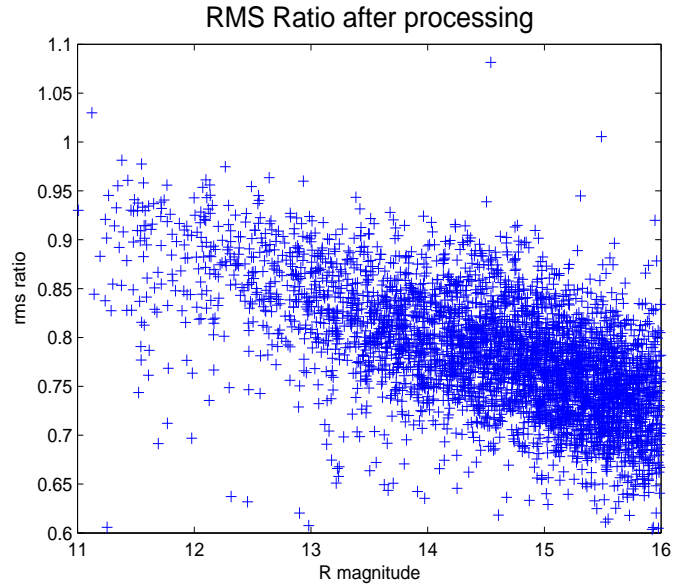}}
\caption{The ratio between the RMS before and after our algorithm (MagZeP and
  exposure-outlier removal and SysRem) was applied.  The figure
  presents 3893 stars observed by the E2 CCD in the first block of
  LRa01, with 1693 exposures}
\label{rmsRatio}
\end{figure*}

%
\section{Discussion}        
\label{discussion}                            
%

We propose here a statistical algorithm to deal with the CoRoT data, as a
complementary process of the regular N2 CoRoT pipeline. The latter
includes only model-based corrections of identified physical effects,
while ours, which is a generalization of a zero-point removal of each
exposure \citep{tmz05, cameron06}, relies only on the collective
effects identified in the data.  We find that the zero point of each
exposure depend on the stellar magnitude. Obviously, other effects
can still be present in the data, and thus in our implementation we
apply SysRem after the MagZeP removal.

The results of applying MagZeP and outlier removal and then applying
SysRem are depicted in Figure~\ref{rmsRatio}, which presents the ratio
between the RMS before and after applying our approach.  We can see
that the improvement is a strong function of the stellar magnitude,
and the {\it averaged} improvement ratio is almost linear with the R
magnitude, and can reach 25\% for the faint stars. Most of the
improvement is due to MagZeP and outlier removal, and adding SysRem
has a minor impact on our approach. Applying SysRem {\it alone} does
achieve less effective clean up.  We find that the improvement
of the light curve is less pronounced at the middle of the LRa01
run in both CCDs. This is because the CoRoT Earth-shadow crossing (see
below) occurs only at the beginning and the end of the run,
generating more pronounced systematic effects.

\begin{figure*}
\centering
\resizebox{\hsize}{!}{\includegraphics{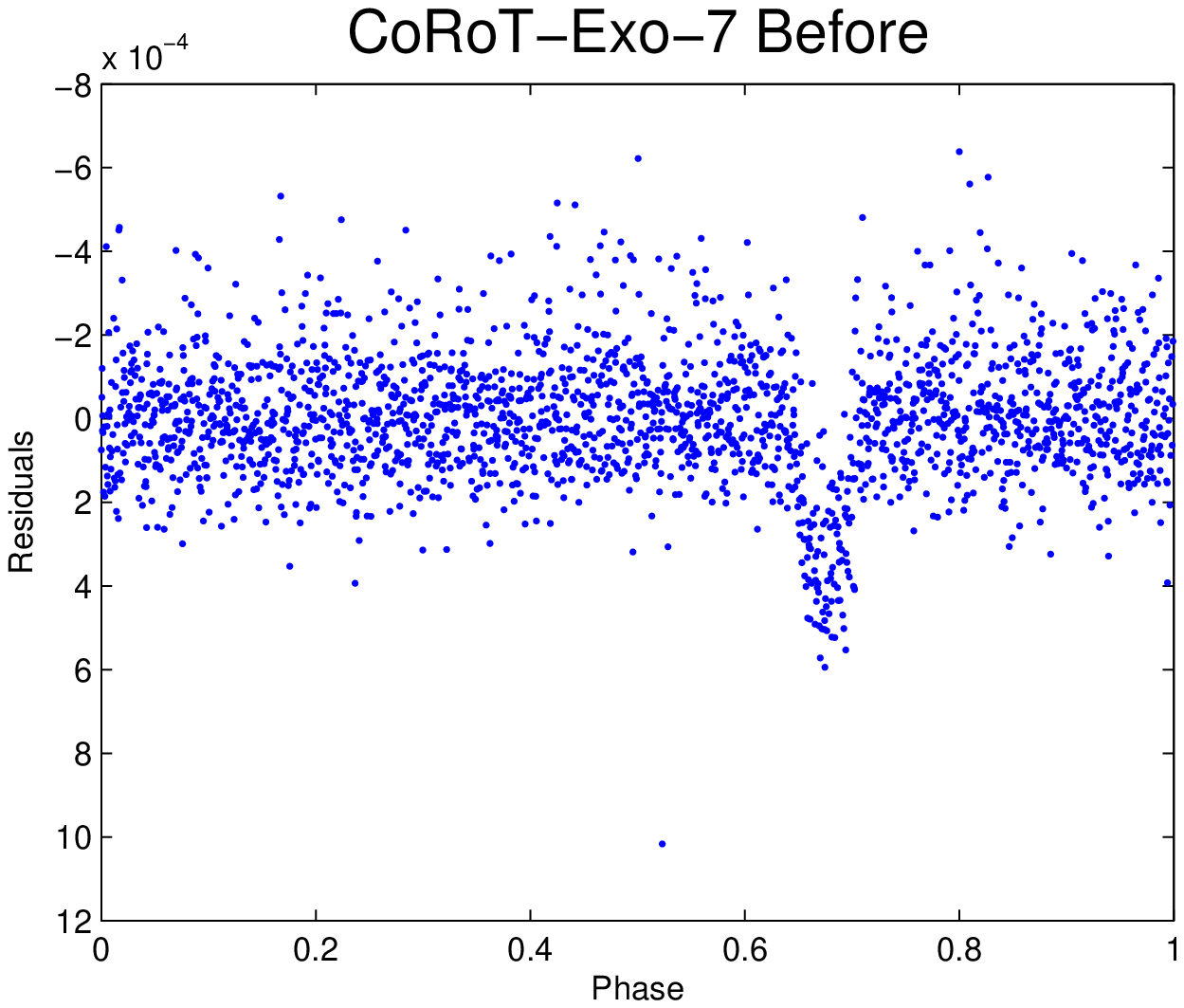}
\includegraphics{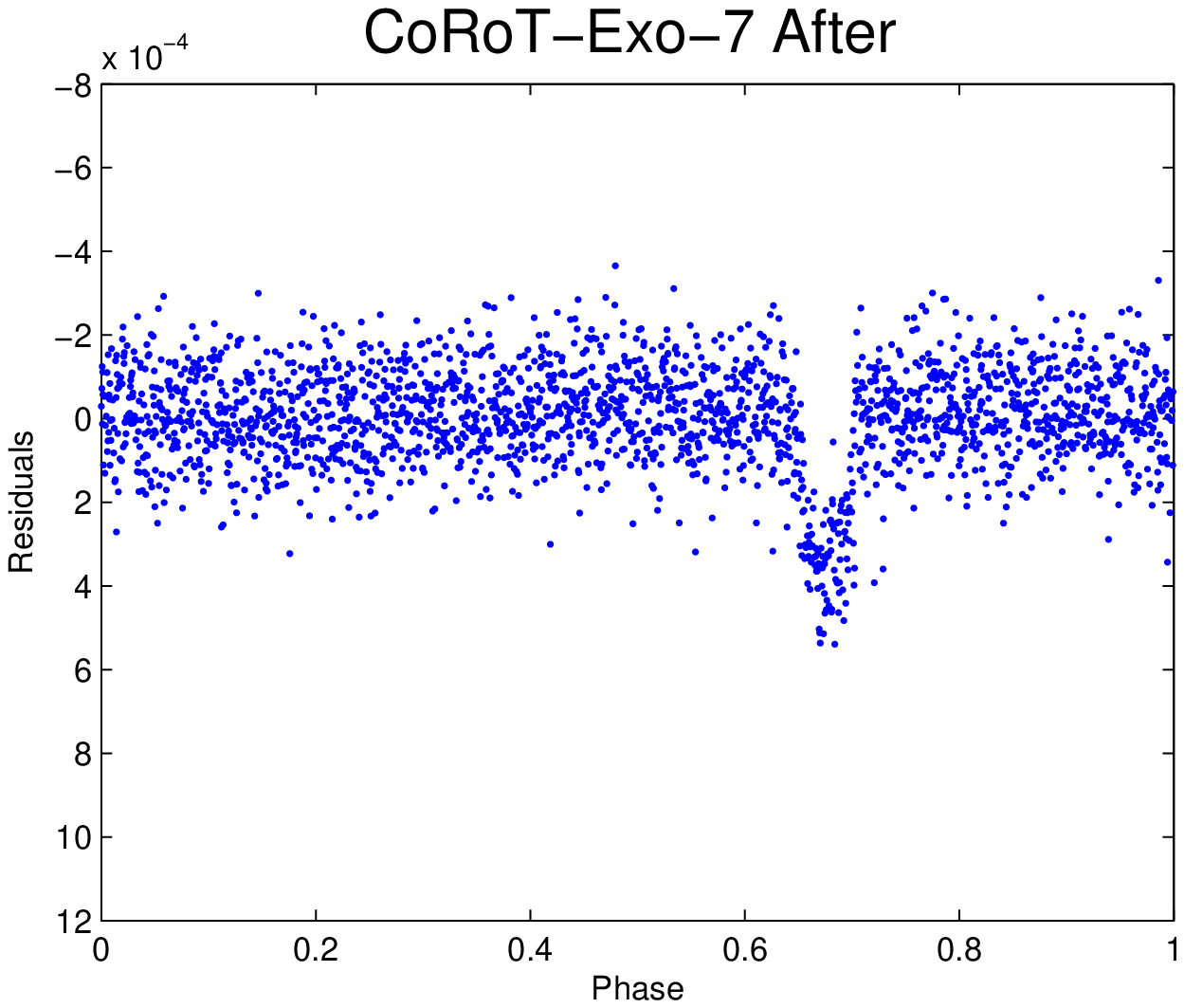}}
\caption{The folded and binned light curve of CoRoT-7, before and
after our analysis.}
\label{exo7}
\end{figure*}

One danger of removing collective effects from a set of light curves
is that the process might attenuate the signal of a possible
transit. To show that this is not the case here we present in
Figure~\ref{exo7} the light curve of CoRoT-7, after removing the
stellar variability and then folding the residuals with the planetary
orbital period of 0.854 days, binning the data such that each bin
presents an average of 10 measurements.  For comparison, the figure
also depicts the light curve before applying our analysis. The signal
is clearer after our analysis.


\begin{figure*}
\centering
\resizebox{10cm}{8cm}
{\includegraphics{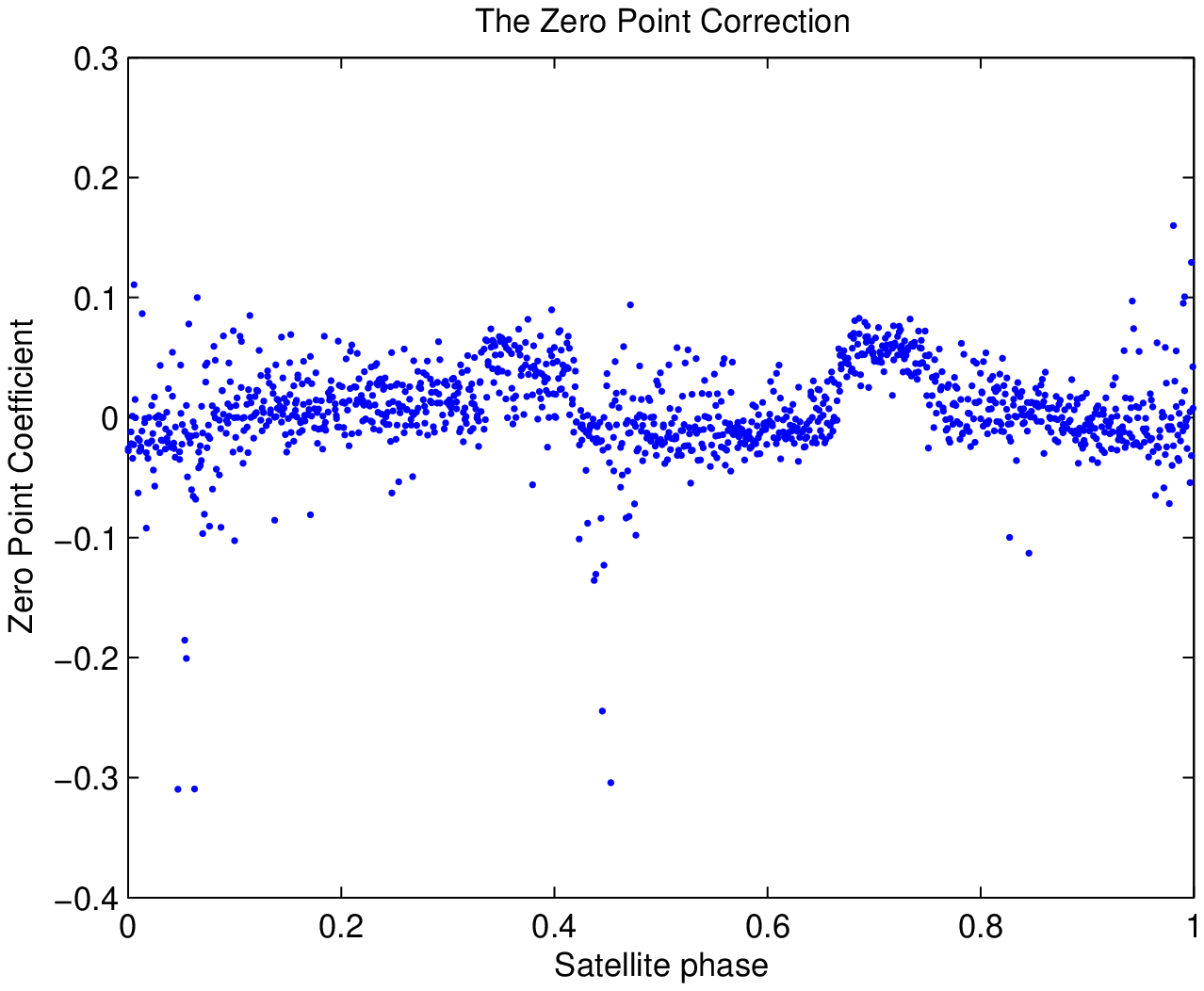}}
\caption{The set of zero-order parameters of all parabolas of the first block,
 $\{a_{0,j},j=1,N\}$, folded with the orbital period of the satellite.
 }
\label{folding1}
\end{figure*}

To look for the source of the MagZeP effect we considered the set of
parameters of the fitted parabolas of all exposures ---
$\{a_{0,j},j=1,N\}$ for example, where $N$ was the number of exposures
included in the block being analysed. This set of parameters reflected
the zero-order brightness removed by our algorithm. We folded
this set of parameters with the orbital period of the satellite, the
results were being plotted in Figure~\ref{folding1}.


The two positive bumps seen in the figure occurred when the satellite
was entering and exiting the Earth shadow, while the negative
outliers, at phase 0.05 and 0.45, coincided with crossing the South
Atlantic Anomaly \citep[see for example][]{auvergne09}. We also
noticed an overall slight curvature over the whole phase, with a small
displacement between the two bumps. The reason for this feature was
not clear.

We wish we had a complete model that accounts for the MagZeP effect,
which is probably associated with an additive and multiplicative
factor of stellar flux. The parabolic fit we use is
only an approximation to the exact function of the MagZeP
effect. However, a detailed analysis of the nature of the effect is
out of the scope of this paper and is therefore deferred for future
work. The goal of this short communication is to point out to the
effect and the impact of its removal.

While trying to improve the algorithm and searching for other
collective effects, we have processed the whole CoRoT N2 white colour
dataset. The cleaned data are now available to the whole CoRoT
community.

%
%

\begin{acknowledgements}
We thank warmly the referee, Ronald Gilliland, for careful reading
the manuscript and suggesting useful and thoughtful comments. T.M. wishes
to acknowledge the support of the Ministry of Science, Culture and
Sport of Israel for France-Israel collaboration in Astrophysics (grant
no. 3-450) and the support by the ISRAEL SCIENCE FOUNDATION (grant
No. 655/07). S.Z. wishes to acknowledge support by the ISRAEL SCIENCE
FOUNDATION -- The Adler Foundation for Space Research (grant
No. 119/07).
\end{acknowledgements}



{}

\end{document}